\documentclass[10pt, paper=a4, UKenglish]{article}
\usepackage{graphicx}
\usepackage{amsmath}
\usepackage{amsfonts}
\usepackage{dsfont}
\usepackage{lineno}
\def\Title#1{\begin{center} {\Large #1 } \end{center}}
\def\Author#1{\begin{center}{ \sc #1} \end{center}}
\def\Address#1{\begin{center}{ \it #1} \end{center}}

\newcommand\pubblock{\rightline{\begin{tabular}{l} Proceedings of the CTD/WIT 2019\\ \pubnumber\\
         \pubdate  \end{tabular}}}

\newenvironment{Abstract}{\begin{quotation} \begin{center} 
             \large ABSTRACT \end{center}\bigskip 
      \begin{center}\begin{large}}{\end{large}\end{center} \end{quotation}}

\newenvironment{Presented}{\begin{quotation} \begin{center} 
             PRESENTED AT\end{center}\bigskip 
      \begin{center}\begin{large}}{\end{large}\end{center} \end{quotation}}

\def\Acknowledgements{\bigskip  \bigskip \begin{center} \begin{large}
      \bf ACKNOWLEDGEMENTS \end{large}\end{center}}

\def\beq{\begin{equation}}
\def\eeq#1{\label{#1}\end{equation}}
\def\eeqn{\end{equation}}

\def\beqa{\begin{eqnarray}}
\def\eeqa#1{\label{#1}\end{eqnarray}}
\def\eeqan{\end{eqnarray}}

\let\bar=\overbar

\def\Dslash{\not{\hbox{\kern-4pt $D$}}}
\def\dslash{\not{\hbox{\kern-2pt $\del$}}}

\def\msb{{\bar{\ssstyle M \kern -1pt S}}}

\newcommand{\pt}{$p_\mathrm{T}$\text{ }}

\usepackage{color}
\usepackage{lineno}
\usepackage{subfig}
\usepackage{hyperref}

\newcommand\pubnumber{PROC-CTD19-029\\ ATL-PHYS-PROC-2019-044}

\newcommand\pubdate{\today}

\def\affiliation{
On behalf of the ATLAS Experiment, \\
SLAC National Accelerator Laboratory, \\
Department of Physics, \\
Stanford University, U.S.A.}

\newcommand{\conference}{Connecting the Dots and Workshop on Intelligent Trackers (CTD/WIT 2019)\\
Instituto de F\'isica Corpuscular (IFIC), Valencia, Spain\\ 
April 2-5, 2019}

\usepackage{fancyhdr}
\definecolor{mygrey}{RGB}{105,105,105}
\begin{document}


\large
\begin{titlepage}
\pubblock

\vfill
\Title{Primary Vertex Selection in VBF Higgs to Invisibles at the HL-LHC with the ATLAS Experiment}
\vfill

\Author{Murtaza Safdari}
\Address{\affiliation}
\vfill

\begin{Abstract}
ATLAS has developed a new approach for primary vertex selection in VBF Higgs invisible events under HL-LHC conditions, exploiting its new forward tracking capabilities, integrating calorimeter and tracking information to mitigate the impact of pileup vertex merging, and introducing a new way to apply pile-up jet suppression methods for the selection of VBF jets. The new algorithm is insensitive to pileup density and improves the average vertex selection efficiency from 86\% to 95\% under tight VBF Higgs event selection cuts.
\end{Abstract}

\vfill

\begin{Presented}
\conference
\end{Presented}
\vfill
\begin{center}
Copyright 2019 CERN for the benefit of the ATLAS Collaboration.
CC-BY-4.0 license.
\end{center}
\end{titlepage}
\def\thefootnote{\fnsymbol{footnote}}
\setcounter{footnote}{0}
%

\normalsize 


\section{Introduction}
\label{intro}

Vertex selection algorithms used to identify the hard-scatter primary vertex among the multiple reconstructed vertices in each bunch crossing at the LHC have thus far relied primarily upon the hardness of the hard-scatter (HS) vertex relative to vertices from pileup (PU). The high PU environment at the HL-LHC \cite{tdr} will, however, introduce major experimental challenges for the correct selection of the HS vertex. In particular, the expected average PU vertex density of around 2 vtx/mm can often lead to the merging of nearby PU vertices resulting in PU vertices with very large summed transverse momentum (\pt\hspace{-4pt}) that can be incorrectly identified as HS by the standard algorithm (Section~\ref{curr}).

This is especially relevant for vector boson fusion (VBF) Higgs invisible final state topologies where the HS process does not have very high visible \pt activity, resulting in a low selection efficiency as a function of PU density when using the standard approach.

A new approach for primary vertex selection in VBF Higgs invisible events under HL-LHC conditions is developed, exploiting the new forward tracking capabilities of the planned ATLAS detector upgrade \cite{tdr}. This new technique integrates tracking information with calorimeter information to mitigate the impact of PU vertex merging, and introduces a new way to apply PU jet suppression for the selection of VBF Higgs events. The new algorithm is insensitive to PU density and improves the average vertex selection efficiency from 86\% to 95\% under tight and unbiased VBF Higgs event selection cuts.

\section{Data Used and Event Selection}
\label{data}

Run 4 simulation data was used ($\langle\mu\rangle=200$, $\sqrt{s}=14$ TeV) for [VBF $H125\rightarrow ZZ\rightarrow 4\nu$] events, with the new Inner Tracker (ITk) geometry layout and reconstruction performance as described in Ref. \cite{tdr} assumed.

Physics event selection ensures that every event has a vertex of interest, setting the theoretical maximum vertex selection efficiency at 100\%. A key challenge in studying primary vertex identification is the fact that physics event selection depends on the choice of the primary vertex. For example, in order to suppress PU, jets are required to have a large fraction of their track \pt pointing to the selected HS primary vertex. This is typically achieved by requiring jet $R_{p_\mathrm{T}}>0.1$, where $R_{p_\mathrm{T}}$ \cite{pu} is defined as:
\begin{equation}
R_{p_\mathrm{T}} = \sum^{\text{tracks}}_{\text{in vertex}} \frac{p_{\mathrm{T}}}{p_{\mathrm{T}(\text{jet})}}\mathds{1}(\Delta R<0.4).
\end{equation}

Here $\Delta R$ is the distance between the track and jet in $\eta-\phi$ space. Since the $R_{p_\mathrm{T}}$ condition requires the definition of an HS vertex, it biases the measurement of the vertex selection efficiency. This is because VBF Higgs events in which the selected vertex is not the true HS primary vertex will not pass the jet selection due to low $R_{p_\mathrm{T}}$ with respect to the incorrect vertex, and will not be considered in the calculation of the efficiency, leading to an overestimation of the vertex selection efficiency.

To overcome this, a new event selection is defined that is unbiased with respect to the choice of the vertex. The idea is to first associate all physics objects (jets in the case of VBF Higgs invisibles) to vertices, and then apply the VBF Higgs event selection to each vertex separately; if at least one vertex contains 2 jets passing the VBF jet criteria, the event is considered for analysis. This ensures that all suitable VBF Higgs events will be considered for the calculation of the vertex selection efficiency, regardless of which vertex was selected by the event selection algorithm, leading to a vertex-unbiased event selection.

Jets are associated to the vertices that maximize $R_{p_\mathrm{T}}$, after which the jets attached to individual vertices are tested for the VBF jet criteria.

\section{Current Approach and its Limitations}
\label{curr}

\begin{figure}[!htb]
  \centering
  \includegraphics[width=1.0\linewidth]{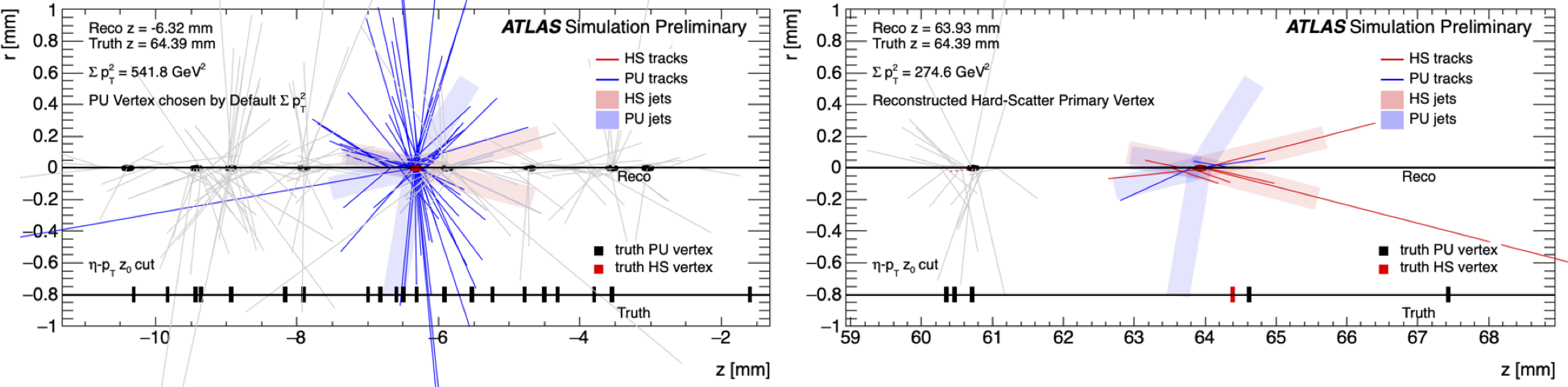}
  \caption{PU vertices with many tracks, as shown in the left r-z event display, dominate the softer reconstructed HS Primary Vertex (in the same Event), on the right, when using sumPT. The left plot is centered (red ellipse) on the vertex incorrectly chosen by sumPT, the right plot is centered on the reconstructed HS Primary Vertex. All the jets with $p_\mathrm{T}>30$ GeV are drawn in the r-z plane on top of the vertex of interest. Also shown in the lower half of the plots is the truth Z axis with the truth HS and PU vertex positions marked.}
  \label{fig:sumpted}
\end{figure}

The current approach to finding the HS vertex in an event is to find the vertex with the largest scalar summed squared track transverse momentum, 
\begin{equation}
\text{sumPT}=\sum_{\text{tracks}} p^2_{\mathrm{T}}.
\end{equation}
One limitation of this method is that, under high luminosity conditions, it becomes vulnerable to selecting high \pt vertices resulting from the merging of nearby PU vertices. Nearby PU vertices can be merged when the typical distance between pileup vertices is comparable or smaller to the vertex resolution. An example where the sumPT algorithm incorrectly selects one of such merged vertices is shown in Figure~\ref{fig:sumpted}. This effect is further illustrated in Figure~\ref{fig:esecomp}, where the sumPT vertex selection efficiency shows a downward trend with increasing average local PU vertex density. This consequence of merged PU vertices is exacerbated for VBF Higgs to invisibles vertices with their modest visible \pt.

\section{New Technique}
\label{new}
The key difference between the selected PU vertex and the correct HS vertex in Figure~\ref{fig:sumpted} is that in the case of the HS vertex, the tracks are correlated with the high \pt jets in the event whereas the tracks within the PU merged vertex are not strongly correlated in angles, because they originate from independent, close-by, interactions. Therefore there is room for improvement by resolving the substructure associated with vertices, and grooming the tracks attached to the vertices such that tracks correlated to HS processes are selected preferentially. 

Based on this insight, we proposed a new algorithm for vertex selection in VBF Higgs invisible events at HL-LHC that makes use of the track-jet correlations within each vertex, based on maximizing an augmented version of the standard sumPT algorithm that projects tracks onto hard jets,
\begin{equation}
\text{sumPTw} = \sum_{\text{tracks}} p_{\mathrm{T}_w} = \sum_{\text{tracks}} p_{\mathrm{T}}^2 \, p_{\mathrm{T}(\text{closest jet})}^2 \frac{1}{\Delta R}\mathds{1}(\Delta R<0.8) \mathds{1}(p_{\mathrm{T}(\text{jet})}>p_{\mathrm{T}(\text{threshold})}).
\end{equation}

Here $p_{\mathrm{T}(\text{threshold})}=30$ GeV is used, and $\Delta R$ is the distance between a track and the closest jet with sufficient \pt in $\eta-\phi$ space. This projects tracks onto jets and adds the weighted $p_{\mathrm{T}_w}$ as defined above. Thus sumPTw is a bolstered measure of the total vertex sumPT in jets. Since merged PU vertices are made of 2 or more independent interactions, their tracks are not correlated in $\eta-\phi$ space with the HS interaction. Therefore by projecting their tracks onto hard jet axis directions, their large sumPT is significantly reduced. 
%

\begin{figure}[!htb]
  \centering
  \includegraphics[width=0.6\linewidth]{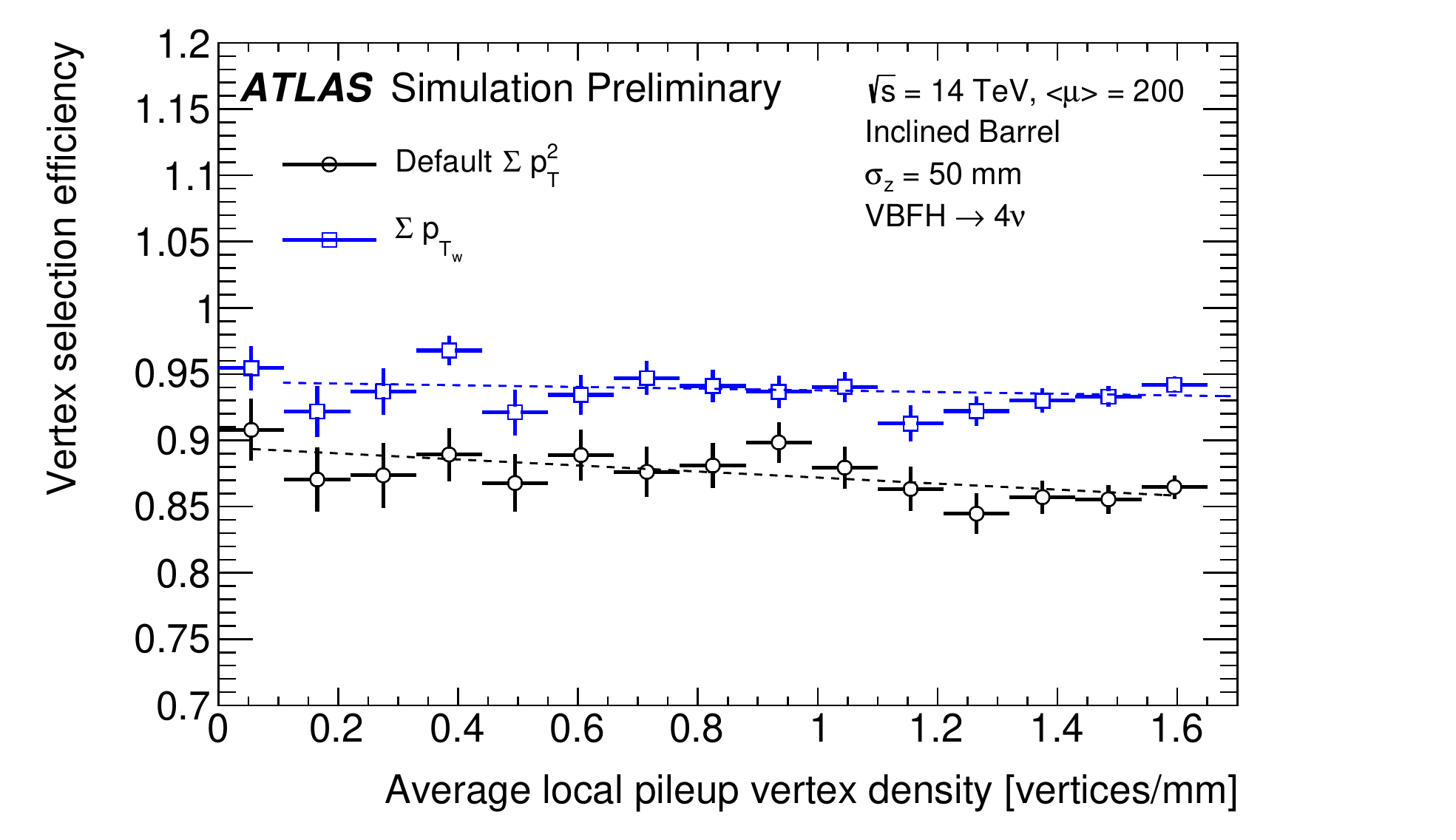}
  \caption{Vertex selection efficiency as a function of the Average local PU vertex density at the reconstructed HS Primary Vertex in the event. Default sumPT selection method (circles) is compared with proposed sumPTw selection (squares). sumPTw modestly flattens the PU vertex density dependence of vertex selection efficiency. Error bars are statistical errors.}
  \label{fig:esecomp}
\end{figure}

Figure~\ref{fig:esecomp} shows the new sumPTw algorithm (square) performance versus the standard sumPT algorithm (circle). sumPTw modestly flattens the PU vertex density dependence of vertex selection efficiency. This demonstrates the robustness of sumPTw in the face of merged PU vertices at high PU vertex densities. The new algorithm improves the average Vertex Selection Efficiency from 86\% to 95\%, a 10\% relative improvement on the current approach under the tight event selection criteria as described in Section~\ref{data}.

\section{Experimental Challenges and Future Improvements}
\label{limits}

The new vertex selection algorithm effectively mitigates the effects of PU vertex merging that would limit the performance of selection in VBF Higgs invisible events. Three experimental challenges are identified that keep the vertex selection efficiency of the new selection technique below 100\% (Figure~\ref{fig:esecomp}); vertex splitting, jets outside detector acceptance, and presence of Hard QCD PU interactions in the event.

Vertex splitting: In events where too many HS tracks have been mis-associated to nearby PU vertices, one finds that the hard PU vertices are selected instead of the HS Primary Vertex. This could manifest in two ways; since the HS vertex has reduced sumPTw, either a random hard PU vertex with large enough sumPTw gets selected, or a PU vertex near the HS Primary Vertex acquires too many tracks pointing along hard jets and gets selected.

Jets outside acceptance: In events where HS jets are outside the detector acceptance the HS Primary Vertex loses out to PU vertices. This is because there are no reconstructed tracks associated to these jets, and thus sumPTw can do nothing to help the Primary Vertices in these events.

\begin{figure}[!htb]
  \centering
  \includegraphics[width=0.6\linewidth]{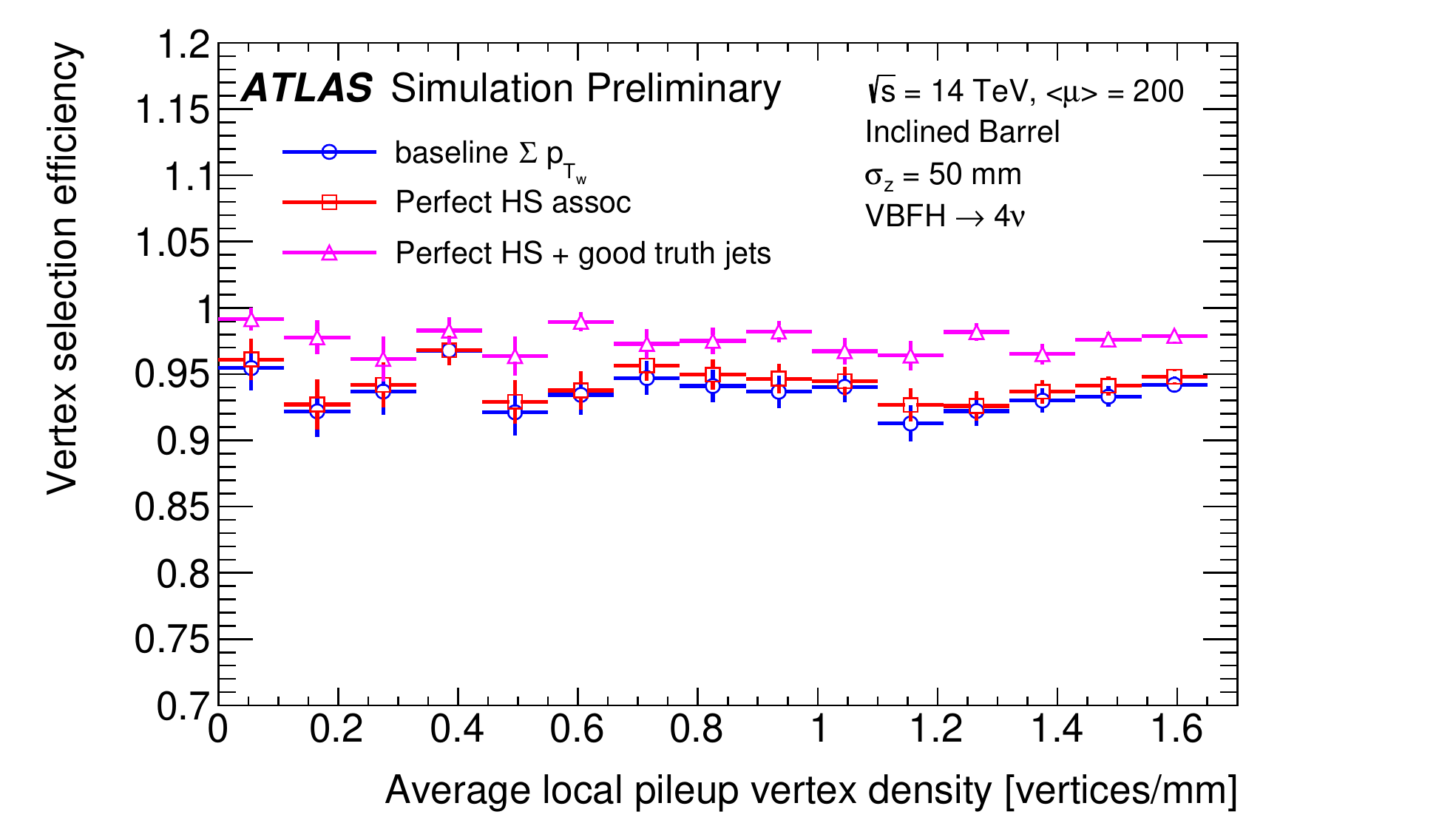}
  \caption{Vertex selection efficiency as a function of the Average local PU vertex density at the reconstructed HS Primary Vertex in the event. The baseline sumPTw performance (circle) is compared with the performance of the sumPTw selection on events where, (square) HS tracks have been reattached to the reconstructed HS Primary Vertex using truth information, and (triangle) Only events with $>2$ truth jets with (truth $p_\mathrm{T}$)$> 30$ GeV \& (truth $|\eta|$)$<4$ are considered. Error bars are statistical errors.}
  \label{fig:esesumptw}
\end{figure}

One can quantify the effect of these two cases, by mitigating vertex splitting by reattaching all the HS tracks to the Primary Vertex using truth information, and by limiting to events with HS jets that are within detector acceptance. The performance with these cases mitigated is shown in Figure~\ref{fig:esesumptw}.

\begin{figure}[!htb]
  \centering
  \includegraphics[width=1.0\linewidth]{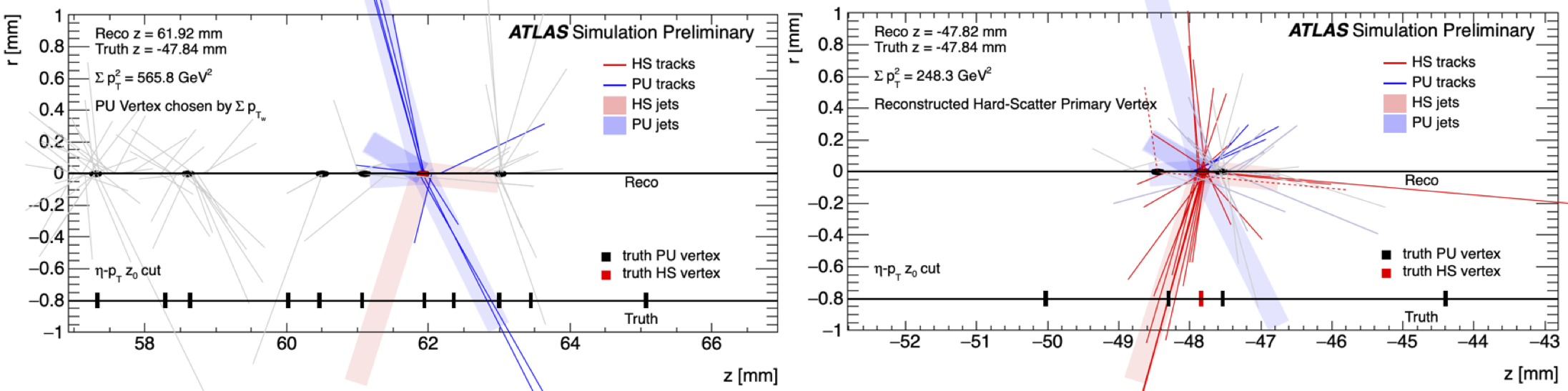}
  \caption{r-z Event display highlighting presence of a Hard QCD PU vertex that gets chosen by sumPTw, instead of the reconstructed HS Primary Vertex. The left plot is centered on the vertex incorrectly chosen by sumPTw (sumPTw=1.2e8), the right plot is centered on the reconstructed HS Primary Vertex (sumPTw=6.7e6). All the jets with $p_\mathrm{T}>30$ GeV are drawn in the r-z plane on top of the vertex of interest. Lower half of the plots show the truth Z axis with the truth HS and PU vertex positions marked.}
  \label{fig:sumptwqcded}
\end{figure}

Hard QCD PU: The remaining cases where the new algorithm runs short is illustrated in Figure~\ref{fig:sumptwqcded}. This is where the event has Hard QCD PU interactions, along with the VBF Higgs event of interest. As a result, the PU vertices have very hard PU tracks pointing along very hard PU jets and thus get selected by sumPTw.

This final limitation is a fundamental shortcoming of topology agnostic vertex selection algorithms; They can't distinguish between different topologies. For this reason sumPTw can pick a Hard QCD PU vertex instead of the VBF Higgs HS vertex, even though there was no malfunction in the vertex selection algorithm. Consequently this paves the way for a more sophisticated vertex selection algorithm, one which extends the innovations presented here and combines them with topology discriminating properties. Such an algorithm would then be able to tell QCD PU apart from HS vertices, despite competing summed scalar transverse momentum, etc.
 
\section{Conclusions}

ATLAS has developed a new approach for primary vertex selection that integrates calorimeter and tracking information to mitigate the impact of PU vertex merging. The new algorithm is insensitive to PU density and improves the average Vertex Selection Efficiency from 86\% to 95\%. 

A possible avenue for improvement upon the new algorithm proposed here involves developing a topology exploiting selection algorithm which can handle Hard QCD PU interactions in the event.

\vspace{-12pt}
\Acknowledgements
We are grateful to Graham Richard Lee, Valentina Cairo, Matthias Danninger, and Nora Emilia Pettersson for fruitful discussions and help with this project.



\end{document}